\documentclass[journal]{IEEEtran}

\usepackage{amssymb}
\usepackage{amsmath}
\usepackage{pifont}
\usepackage{graphics}
\usepackage{cite}
\usepackage{epsf}
\usepackage{epsfig}

\newcommand{\Cu}{\ensuremath{\mathrm{Cu}}}
\newcommand{\Ni}{\ensuremath{\mathrm{Ni}}}
\newcommand{\Nb}{\ensuremath{\mathrm{Nb}}}
\newcommand{\Al}{\ensuremath{\mathrm{Al}}}

\renewcommand{\O}{\ensuremath{\mathrm{O}}}
\newcommand{\F}{\ensuremath{\mathrm{F}}}
\renewcommand{\S}{\ensuremath{\mathrm{S}}}
\newcommand{\Ar}{\ensuremath{\mathrm{Ar}}}

\begin{document}
\title{Josephson junctions with centered step and local variation of critical current density}

\author{M. Weides
\thanks{Manuscript received 26 August 2008. This work was supported by the  DFG project WE 4359/1-1.}
\thanks{M. Weides is with the Institute of Solid State Research and JARA- Fundamentals of Future Information Technology, Research Centre, J\"ulich, 52425 J\"ulich, Germany (e-mail: m.weides@fz-juelich.de).}}

\maketitle


\begin{abstract}
Super\-conductor-insulator-ferromagnet-su\-per\-con\-duc\-tor (SIFS) Josephson tunnel junctions based on $\Nb/\Al_2\O_3/\Ni\Cu/\Nb$ stacks with a thickness step in the metallic $\Ni\Cu$ interlayer were fabricated. The step height of a few $0.1\;\rm{nm}$ was defined by optical lithography and controlled etching of both $\Nb$ and $\Ni\Cu$ layers. Experimentally determined junction parameters by current-voltage characteristics and Fraunhofer pattern indicate a uniform $\Ni\Cu$ thickness and similar interface transparencies for etched and non-etched parts. The critical current diffraction pattern was calculated and measured for stepped junctions having the same ground phase difference but different critical current densities in both halves. The measured data show a good agreement with simulations.
\end{abstract}


%

\section{Introduction}

\PARstart{T}{he} \emph{Josephson junction} (JJ) is considered as the work horse in superconducting electronics. They are based on two weakly coupled superconducting electrodes via a constriction, e.g., made up by a normal metal or a tunnel barrier and are routinely applied in ultra-high sensitive SQUID (Superconducting Quantum Interference Devices) magnetometers, or the voltage standard \cite{BuckelKleiner2004Superconductivity}. Especially $\Nb/\Al_2\O_3/\Nb$ tunnel junctions attract considerable interest as the fabrication of high density $\Nb$-based Josephson circuits with promising small parameters spreads is possible. With the advent of high-quality magnetic tunnel junctions one decade ago, new so far unexplored devices are now under development which combine both fabrication techniques by advanced multilayers of superconducting (S), insulating (I) and magnetic (F) materials. These superconducting spintronic devices were in the focus of recent research activities, like so called $0$--$\pi$ Josephson junctions \cite{WeidesFractVortex,PfeifferPRB08} where the type of coupling ($0$ or $\pi$) is related to the local thickness of the stepped F-layer barrier in the junction. In this work stepped JJs with variation of critical current densities, but same coupling, are discussed.\par Generally, for a variety of JJs a non-uniform critical current density $j_c$ is desirable, e.g., for tunable superconducting resonators, toy systems for magnetic flux pinning or magnetic-field driven electronic switches being similar to SQUIDs.\par
The first considerations \cite{Russo1978NonuniformJc} of non-uniform $j_c$'s were caused by technological drawbacks leading to a variation of the effective barrier thickness by either fabrication \cite{Schwidtal1969} or illumination of light-sensitive barriers \cite{Barone1977PhysStat}. JJs with periodic spatially modulations of $j_c$ were intensively studied regarding the pinning of fluxons \cite{McLaughlinScottPRB1978,Vystavkin1988,MaloUstinovJAP1990}, the spectrum of electromagnetic waves \cite{Lazarides2005,FistulPRB1999} or their magnetic field dependencies \cite{LazaridesPeriodicDefects2003}. Experimentally, a modulation of $j_c$ was done by lithographic insertion of defects such as i) insulation stripes ($j_c=0$) \cite{GolubovUstinovPLA1988,ItzlerTinkhamPRB95}, microshorts ($j_c$ increased) or microresistors ($j_c$ decreased).\\The properties of JJs depend on geometrical (width, length, thickness) and the physical (dielectric constant $\epsilon$, specific resistance $\rho$, magnetic thickness $\Lambda$ and $j_c$) parameters. When tailoring $j_c$ all other parameters should be unchanged to facilitate calculations and avoid further inhomogeneities in the system. These conventional methods for changing $j_c$ necessarily modify either $\epsilon$ or $\rho$.\par
In this paper a new method is used to gradually modify $j_c$ in one half of the junction. The fabrication technology presented (see Fig. \ref{SFpatterning}) permits the controlled change of only the interlayer thicknesses $d_1$ and $d_2=d_1+\Delta d$, i.e., the local $j_c$, while keeping both $\epsilon$ and $\rho$ constant. The magnetic field dependence of the critical current is measured for several stepped junctions and compared to simulations.

\section{Experiment}

\begin{figure}[tb]
\begin{center}
\includegraphics[width=8.6cm]{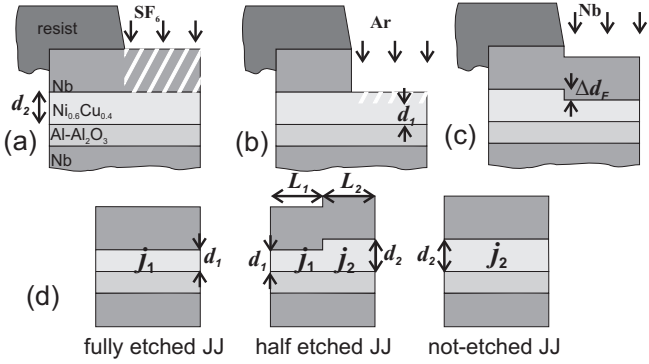}
\caption{\label{SFpatterning} The complete SIFS stack was protected in part by photoresist. (a) reactive etching of
$\Nb$ with $\S\F_6$ down to $\Ni\Cu$ layer, (b) ion-etching of $\Ni\Cu$ to increase the local $j_c$, (c) in situ deposition of cap $\Nb$ layer and (d) the cross-section of the two reference junctions (having fully and not-etched interlayer) and the stepped (half-etched) junction.}
\end{center}
\end{figure}

The deposition and patterning of the stepped junctions was performed by a four level photolithographic mask procedure \cite{WeidesFabricationJJPhysicaC,WeidesSteppedJJ}. The SIFS stack were deposited by a magnetron sputter system.  $\Nb$ and $\Ni\Cu$ were statically deposited, whereas $\Al$ was deposited during sample rotation and at much lower deposition rates to obtain very homogeneous and uniform films.\\
After the deposition of the $\Nb$ as cap layer and subsequent lift-off the complete SIFS stack, without steps in F-layer yet, was obtained.
The part of the JJ that was supposed to have a larger thickness $d_2\approx5\textrm{-}7\:\rm{nm}$ was protected by photoresist, see Fig.~\ref{SFpatterning}.
It was shown that $\S\F_6$ reactive ion etching provides an excellent chemistry for low-voltage anisotropic etching of $\Nb$ with high selectivity towards other materials\cite{SF6EtchingLichtenberger1993} and the photoresist. The inert $\S\F_6$ dissociated in a RF-plasma and the fluor reacted with niobium
$5\F+\Nb\rightarrow\Nb\F_5$. The volatile $\Nb\F_5$ was pumped out of the etching-chamber.
When $\S\F_6$ was used as process gas all non-metallic etching products such as fluorides and sulfides from the top-layer of the $\Ni\Cu$-layer had to be removed by subsequent argon etching. \par
The patterning process of the step is depicted in Fig.~\ref{SFpatterning} (a)--(c). The key points were a) \emph{selective reactive etching} of $\Nb$, b) \emph{argon etching} of $\Ni\Cu$ to define $d_1=d_2-\Delta d_F$ and c) subsequent \emph{in situ} deposition of $\Nb$.\par
The $\Nb$ cap layer was removed by reactive dry etching using $\S\F_6$. A few $0.1\;\rm{nm}$ of $\Ni\Cu$ were $\Ar$ ion etched at a very low rate to avoid any damaging of the $\Ni\Cu$ film under the surface and to keep a good control over the step height, see etching rates in table \ref{sputter}. The etching was stopped when the F-layer thickness was reduced down to the thickness $d_1$ and subsequently $\Nb$ was deposited as cap-layer. The complete etching and subsequent $\Nb$ deposition was done in-situ. The chip contained stacks with the new $\Ni\Cu$ thicknesses $d_1$ (uniformly etched), $d_2$ (non-etched) and with step in the F-layer thickness  $\Delta d_F$ ranging from $d_1$ to $d_2$ as depicted in Fig. \ref{SFpatterning} d).\\
\begin{table}[tb]
\caption{Etching parameters of $\Nb$ and $\Ni\Cu$. The rates were determined by profiler measurements.\label{sputter}}
\begin{tabular}{ccccc}
  \hline
  \hline
   & partial pressure  & power density  & etching rate & $\quad$ \\
  &[$\rm{mbar}$]&[$\rm{W/cm^2}$] & [$\rm{nm/s}$]& \\
  \hline

  $\S\F_6$ on \Nb & $15\cdot10^{-3}$ & $0.6$ &$\sim1$ \\
  $\S\F_6$ on \Ni\Cu & $15\cdot10^{-3}$ & $0.6$ & $<$0.001 \\
  $\Ar$ on \Ni\Cu & $5\cdot10^{-3}$ & $0.6$ &$\sim0.01$ & \\
   \hline
  \hline
\end{tabular}
\end{table}
The junction mesas were defined by aligning the photo mask on the optically visible step-terraces, followed by $\Ar$ ion-beam etching of the upper $\Nb$, $\Ni\Cu$ and $\Al$ layers. The length of both junctions halves $L_1$ and $L_{2}$ are within the lithographic alignment accuracy of $1\;\rm{\mu m}$. The etching was stopped after the complete etching of the $\Al_2\O_3$ tunnel barrier. Afterwards the mesas were insulated by SNEAP (Selective Niobium Etching and Anodization Process) \cite{Gurvitch82NbAlONb}. In the last photolithographic step the wiring layer was defined. After a short argon etch to remove the contact resistance the thick $\Nb$ wiring was deposited.

\section{Results and Discussion}
Three different types of junctions are discussed: being either fully etched or not-etched (so-called reference JJs) and half etched (stepped JJs) (Fig. \ref{SFpatterning} d). From the current-voltage characteristic (IVC) and magnetic field dependence of critical current $I_c(H)$ of the reference JJs one can estimate parameters for the stepped junction, such as the ratio of symmetry $\Delta=j_2/j_1$, where $j_1=j_c(d_1)$ and $j_2=j_c(d_2)$, and the quality of the etched and non-etched parts.\par
The uniformity of the supercurrent transport in a Josephson junction can be judged qualitatively from the $I_c(H)$ pattern. The magnetic field $H$ was applied in-plane and along one junction axis. The magnetic diffraction pattern depends in a complex way on the current distribution over the junction area \cite{BaronePaterno} and the effective junction length. The ideal pattern of a short ($L_1+L_2\leq\lambda_J$) JJ with Josephson penetration depth $\lambda_J$ is symmetric with respect to both polarities of the critical
current and the magnetic field and has completely vanishing  $I_c$ at the minima. Asymmetry, irregularity or current offsets in $I_c(H)$ indicate a non-uniform current transport over the interlayers. If the JJs is flux-free, this non-uniformity can be located in both the insulating and ferromagnetic layers as well as at the interfaces.\\Transport measurements were made in a liquid He dip probe using low-noise home made electronics and room-temperature voltage amplifier. The critical current was determined by a voltage criteria of $3\:\rm{\mu V}$.\par

\subsection{Uniformly etched junction}
\begin{figure}[tb]
\begin{center}
  \includegraphics[width=8.6cm]{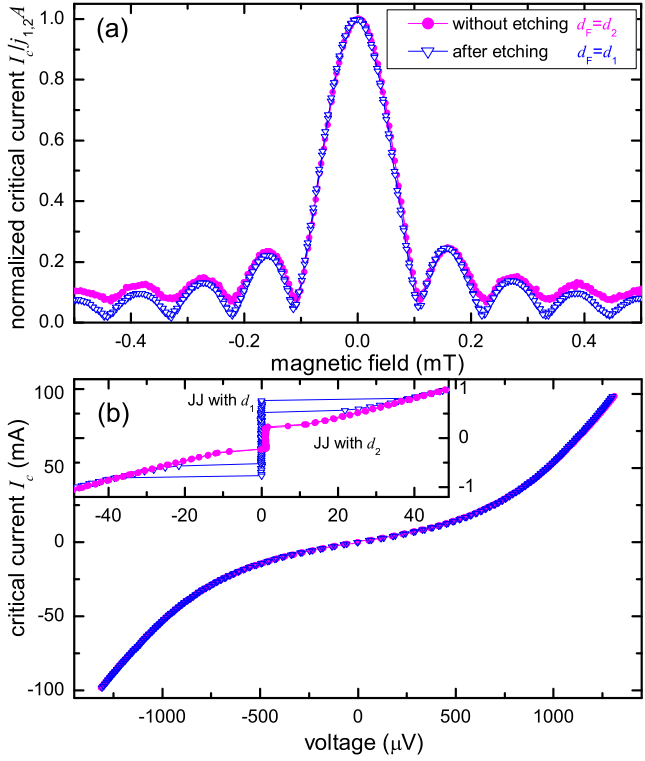}
  \caption{(color online) a) $I_c(H)$ of etched (circle) and non-etched (triangle) JJs. $I_c$ was normalized to the maximum value $j_1\cdot A$ or $j_2 \cdot A$. b) IVCs for large and small (inset) bias current ranges, measured at zero magnetic field. In b) the IVC of not-etched JJs is nearly completely covered by the IVC of fully etched JJ. The junction lengths are $100\:\rm{\mu m}$, both JJs are in the short JJ limit. Measurements were done at $4.2\;\rm{K}$.}
  \label{Fig:IcHIVtogether}
\end{center}
\end{figure}

In this subsection the properties of reference JJs are discussed.
The $I_c(H)$ dependencies for a non-etched junction (triangle) with the F-layer thickness $d_2$ and a uniformly etched junction
(circle) with thickness $d_1=d_2-\Delta d_F$ are depicted in Fig.~\ref{Fig:IcHIVtogether} a). Their $I_c(H)$ pattern are normalized to the maximum value $j_1\cdot A$ or $j_2 \cdot A$, respectively, with junction area $A$. The larger offset of the non-etched JJs is due to their lower absolute critical current. Fig.~\ref{Fig:IcHIVtogether} b) depicts the IVCs for large and small (inset) ranges of bias current. As the electric transport is in the dirty limit \cite{VasenkoPRB}, $j_c$ scales exponentially with the variation of $d_F$.  The polycrystalline structure of room-temperature sputtered layers and the very low etching rate of $\Ni\Cu$ led to a good control over $\Delta d_F$. However, one has to keep in mind that the local variation of F-layer thickness might exceed this value, and $d_1$, $d_2$ and $\Delta d_F$ are just the mean thicknesses seen by the transport current. Besides the difference in $I_c$, the $I_c(H)$ dependence and the IVCs showed no evidence for an inhomogeneous current transport for both samples. The resistance $R$ is nearly independent from $d_F$ as the voltage drop over the tunnel barrier is much larger than the serial resistance of a few nanometer thick metal \cite{WeidesHighQualityJJ}. However, an etching-induced change of transparency at the F/S interface might modify $R$. No change of $R$ is visible in the IVCs of both JJs in Fig.~\ref{Fig:IcHIVtogether} b), apart from the change in $I_c$. A change of capacitance $C$ requires a change of $R$. Both $R$ and $C$, i.e., $\rho$ and $\epsilon$, are determined by the dielectric tunnel barrier, thus the only difference between both junctions is the local $j_c$.  The larger $I_c$, but same resistance $R$ and capacitance $C$ led to a slightly hysteretic IVC of the etched sample, as the width of hysteresis is determined by the McCumber parameter $\beta_c \propto I_c^2 R C$.\par

\subsection{Half-etched (stepped) junctions}
Stepped JJs with a centered step and same ground state in both halves are treated in this subsection. A theoretical review of the magnetic diffraction pattern for junctions with different $j_1$ and $j_2$ ratios is given and compared with recent measurements.
\subsubsection{Calculated $I_c(h)$ of stepped JJ}
\begin{figure}[tb]
\begin{center}
  \includegraphics[width=8.6cm]{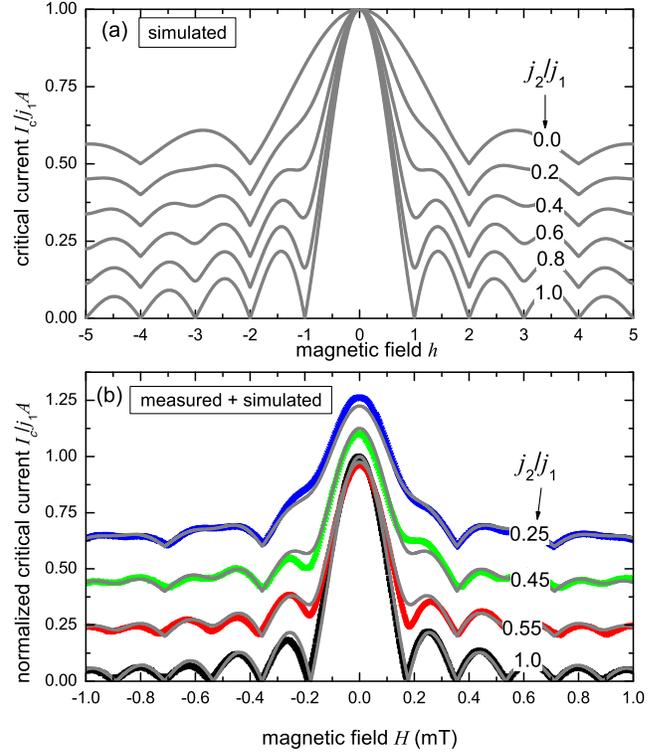}
  \caption{(color online) a) Calculated $I_c(h)$ dependence for various ratios of $\Delta=j_2/j_1$ and centered step in $j_c$ profile. b) Measured $I_c(H)$ of stepped JJs plus calculated $I_c(h)$ (grey line). $I_c$ was normalized to the maximum value $j_1\cdot A$. The junctions are in the short JJ limit and measurements were done at $4.2\;\rm{K}$. Magnetic field is applied in-plane of the junction. The data are shifted along $I_c$-axis.}
  \label{Fig:simuSteJJIcH}
\end{center}
\end{figure}

The magnetic diffraction pattern $I_c(H)$ of a JJ depends on its $j_c$ profile, see Ref.~\cite{BaronePaterno}. The analytic solution for a short stepped junction with centered step ($L_1=L_2=L/2$) and different critical current density $j_1$ and $j_2$ in both halves is given by
\begin{align*}&I_c(h)=w\left[\int\limits_{-L/2}^0\!{j_1\sin{(\phi_0+\frac{hx}{L})}dx}+\int\limits^{L/2}_0\!{j_2\sin{(\phi_0+\frac{hx}{L})}dx}\right]\\
&=A\cdot\frac{j_1\cos{(\phi_0-\frac{h}{2})}-j_{2}\cos{(\phi_0+\frac{h}{2})}+(j_2-j_1)\cos{\phi_0}}{h},
\end{align*}
where $\phi_0$ is an arbitrary initial phase, $h=2\pi\Lambda\mu_0LH/\Phi_0$ the normalized magnetic flux through the junction cross section, $\Lambda$ the magnetic thickness of junction and $w$ the junction width.  The phase-field relation for maximum $I_c$ is reached for
\[\phi_0=\arctan{\left[\frac{\sin{(\frac{h}{2})}\cdot(j_{1}+j_2)}{2\sin^2{\left(\frac{h}{4}\right)}\cdot \left(j_2-j_1\right)}\right].}\] The general analytical form of $I_c(h)$ for multi-step junctions can be found in Ref.\cite{Lazarides2004IcH0PI}.\\
The calculated $I_c(h)$ for various ratios of $\Delta=j_2/j_1$ are depicted in Fig.~\ref{Fig:simuSteJJIcH} a). Characteristic features are the centered maximum peak and the appearance of periodic minima of $I_c(h)$. The depths of the odd-order minima depend on the asymmetry ratio $\Delta$ and decrease for smaller values of $j_2$. $I_c(h)$ is completely vanishing at the even-order minima. The maximum critical current at $I_c(0)$ decreases linearly down to $I_c=0.5\cdot j_1A$ for $j_2=0$. The corresponding $I_c(h)$ pattern becomes that of a junction with half the width and uniform $j_c=j_1$.

\subsubsection{Measured $I_c(H)$ of stepped JJ}
The measured JJs are in short limit, i.e., $L=50\:\rm{\mu m}<\lambda_J$.
In Fig.~\ref{Fig:simuSteJJIcH} b) the measured magnetic diffraction pattern $I_c(H)$ of stepped JJs with different ratio of $\Delta=j_2/j_1$ along with the calculated $I_c(h)$ curves (grey lines) are depicted. The step in interlayer $\Delta d_F$ is carefully varied to trace out various regimes of $I_c(H)$ as function of $\Delta$. $j_1$ and $j_2$ were determined from reference junctions being either fully or not-etched and located close to the stepped junction. Both junctions halves have a phase difference of $\pi$ in the ground state, i.e., the stepped junction was $\pi$ coupled, too. The magnetic field axis $h$ was scaled to fit the first measured minima of $I_c(H)$.\\It can be seen that for smaller $\Delta$ i) the oscillation period changes to the double value by comparing the $I_c(H)$ for $\Delta=1$ and $\Delta=0.25$, ii) the depth of the odd-order minima decreases and iii) the maximum critical current $I_c(0)$ is reduced.\\ The slight asymmetry of some $I_c(H)$ pattern, for example at the first side-maxima of the $\Delta=0.45$ sample, and in consequence the deviation to simulation can be explained by a modification of the magnetic flux penetration due to different magnetic states in both halves \cite{KemmlerPRB0Pi}.\\Recently, it was shown by the author that remanent magnetization of F-layer can lead to strong deviation of the expected $I_c(H)$ pattern \cite{WeidesAnisotropySIFS}. Here, the weak magnet $\Ni\Cu$ was used as interlayer. However, as both halves are in the $\pi$ coupled state the magnetic properties of F-layer can not be neglected. The difference in F-layer thickness for both halves even facilitates some variation of the local magnetic configuration. Nevertheless, the fair agreement of measurement and simulation in Fig.~\ref{Fig:simuSteJJIcH} b) shows the good reliability of the step formation procedure.

\section{Conclusions and Outlook}
As conclusion, SIFS Josephson junctions were fabricated with a well-defined step by local etching of the ferromagnetic interlayer. The etched and not-etched SIFS junctions differ only by F-layer thickness. No inhomogeneities can be seen in the current transport characteristics of the etched junctions. Magnetic field transport measurements on stepped junctions have a good correlation with the simple analytical model.\par As an outlook, the use of a non-magnetic stepped layer avoids the modifications of magnetic cross-section by intrinsic magnetic remanence, and may help to further improve the consistence of $I_c(H)$ measurement with simulation. Replacing the optical lithography with electron beam lithography may enhance the lateral accuracy of the step down to the dimension of e-beam resist. The patterning of stepped JJs allows free lateral placement of well-defined $j_c$'s and/or local coupling regimes within a single junction. JJs with varying $j_c$ and planar phase could be used for devices with special shaped $I_c(H)$ pattern \cite{BaronePaterno}, toy systems for flux pinning or tunable superconducting resonators. The stepped junctions can be realized in $\Nb$ based JJs with any interlayer material, which is chemically stable towards the reactive etching gas.\\The patterning process could be adjusted to all thin film multilayer structure providing that the reactive etching rates of the layer materials differ, e.g., it can be applied to other metallic multilayer systems such as magneto-resistance devices (GMR/TMR elements) where a local variation of magnetic properties may enhance their functionality.

\section*{Acknowledgement}
The author thanks H. Kohlstedt, U. Peralagu and E. Goldobin for fruitful discussions.



\end{document}